\documentclass[a4paper,11pt]{article}
\usepackage{pos}
\usepackage{aas_macros}
\usepackage{amsmath,amssymb,amsfonts}
\usepackage[utf8]{inputenc}

\renewcommand{\L}{\mathcal{L}}

\title{Extending the sample of core-collapse supernovae for searches of axion-like-particle induced gamma-ray bursts with the Fermi LAT}
 \ShortTitle{Searches for ALP induced $\gamma$-ray bursts with the LAT}

\author*[a]{Manuel Meyer}
\author[b]{Tanja Petrushevska}

\affiliation[a]{Institute for Experimental Physics, University of Hamburg,\\
  Luruper Chaussee 149, 22761 Hamburg, Germany}

\affiliation[b]{ Centre for Astrophysics and Cosmology, University of Nova Gorica\\
Vipavska 11c, 5270 Ajdov\v{s}\u{c}ina, Slovenia}

\forColl{\textit{Fermi}-LAT} 

\emailAdd{manuel.meyer@desy.de}
\emailAdd{tanja.petrushevska@ung.si}

\abstract{
During a core-collapse supernova (SN), axion-like particles (ALPs) could be produced through the Primakoff process and subsequently convert into gamma rays in the magnetic field of the Milky Way. Using a sample of well studied extragalactic SNe at optical wavelengths, we estimate the time of the core collapse and search for a coincident gamma-ray burst with the Fermi Large Area Telescope (LAT). Under the assumption that at least one SN was contained within the LAT field of view, we exclude photon-ALP couplings within a factor of $\sim$5 of previous limits from SN1987A. With the increasing number of SNe observed with optical surveys, our results demonstrate the potential to probe ALP dark matter with combined optical and gamma-ray observations. 
We also provide preliminary results for the estimation of explosion times of 15 close-by SNe observed recently with ZTF.
Our findings show that the explosion time can be estimated within one day (statistical uncertainty only) making them promising targets for a follow-up LAT analysis.
}

\FullConference{37$^{\rm{th}}$ International Cosmic Ray Conference (ICRC 2021)\\
		July 12th -- 23rd, 2021\\
		Online -- Berlin, Germany}

\begin{document}
\maketitle

\section{Introduction}
Axions, and more generally axion-like particles (ALPs), are predicted in numerous extensions of the standard model of particle physics~\cite{jaeckel2010}.
If sufficiently light, these particles are also candidates to explain the dark matter density of the Universe~\cite[e.g.,][]{arias2012}.
A promising avenue to search for ALPs is their predicted coupling to photons which is described through the Lagrangian density $\mathcal{L}_{a\gamma} = g_{a\gamma} \mathbf{E}\cdot\mathbf{B}a$, where $g_{a\gamma}$ is the coupling constant between photons and ALPs, $\mathbf{E}$ is the electric field of the photon, $\mathbf{B}$ denotes an external magnetic field, and $a$ is the ALP field strength~\cite{raffelt1988}. 
This coupling leads to the possibility that ALPs are produced through the Primakoff process, i.e., the conversion of $\gamma$-rays in the electrostatic fields of ions in the dense plasma of a core-collapse supernova (SN)~\cite[e.g.,][]{2015JCAP...02..006P}.
Similar to the neutrinos produced during the core collapse, the produced ALPs would quickly escape the core and could convert back to $\gamma$ rays in the magnetic field of the Milky Way. 
As a result, one expects a short ALP-induced $\gamma$-ray burst that arrives in coincedence with the produced neutrinos. 
Assuming only the Primakoff process, the authors of Ref.~\cite{2015JCAP...02..006P} found that 
for a progenitor with $10\,M_\odot$, where $M_\odot$ denotes the mass of the sun, the ALP (and consequently the $\gamma$-ray) spectrum should peak around 60\,MeV and the burst should last for about 10\,s. 

The non-observation of a $\gamma$-ray burst from SN1987A with the Solar Maximum Mission in coincidence with the neutrino detection led to stringent limits on the photon-ALP coupling, $g_{a\gamma} \lesssim 5.3\times10^{-12}\,\mathrm{GeV}^{-1}$ for ALP masses $m_a \lesssim 10^{-10}$\,eV~\cite{2015JCAP...02..006P}.\footnote{See, however, Ref.~\cite{2017arXiv171206205B} for a criticism on the assumptions entering the calculations of the ALP production rate in SNe.} 
Such constraints could be improved by more than an order of magnitude if a Galactic SN occurred in the field of view (FOV) of the Large Area Telescope (LAT) on board the \emph{Fermi} satellite~\cite{2017PhRvL.118a1103M}. 
The LAT is a pair-conversion telescope which detects $\gamma$~rays in the energy between 20\,MeV and beyond 300\,GeV. 
However, with a rate of roughly 3 SNe per century~\cite[e.g.][]{adams2013}, the detection probability of such an event with the LAT are marginal. 

Another possibility is to search for the burst signal from a sample of close-by extragalactic SNe discovered with optical surveys. 
However, current neutrino telescopes lack the sensitivity to detect a neutrino signal from these more distant explosions~\cite{kistler2011}.
Alternatively, one can try to estimate the time of the core collapse using the observations of the optical surveys.
We have conducted such a search using $\gamma$-ray observations of the LAT~\cite{2020PhRvL.124w1101M} and we summarize our findings in Section~\ref{sec:previous-results}. 
In Section~\ref{sec:ztf-sample}, we provide a preliminary set of SNe recently observed with the Zwicky Transient Factory (ZTF)~\cite{2019PASP..131f8003B}, which can be used in a future LAT analysis to extend the SN sample. 
Using the publicly available \textsc{MOSFiT} code, we fit the optical light curves of these new SNe in order to estimate the explosion time. 
Such an estimate is necessary to conduct a search for a $\gamma$-ray signal in the correct time interval. 
We conclude in Section \ref{sec:conclusions}.

\section{Searches for ALP-induced $\gamma$-ray bursts from extragalactic SNe with the LAT}
\label{sec:previous-results}

In Refs.~\cite{2020PhRvL.124w1101M,2020PhRvL.125k9901M}, we selected a sample of 20 extragalactic SNe from the open supernova catalog~\cite{2017ApJ...835...64G} with redshifts in the range $0.003 < z < 0.059$ to search for a potential $\gamma$-ray signal.
We restricted the sample to SNe of type Ib and Ic.
This is because the progenitors are massive blue supergiants for which the time delay between the core collapse and the shock breakout (SBO) is believed to be shorter (less than one day) than for red super giants~\cite{Kistler:2012as}.
To estimate the time of the SBO, we first used the \textsc{MOSFiT} code~\cite{2018ApJS..236....6G} to fit the optical light curves with different models for the evolution of the SN emission. Three different engines were assumed to power the SN, namely the radioactive decay of Nickel and Cobalt (NiCo), the NiCo engine together with an additional SBO component, as well as a generic exponential rise and power-law decay (exppow); see the supplemental material of Ref.~\cite{2020PhRvL.124w1101M} and Ref.~\cite{2017ApJ...849...70V} for a summary of these models. 
All three models generally resulted in a good description of the optical light curves. 
The posterior distribution of the explosion time, $t_\mathrm{exp}$, which is a free parameter in all engine models, was extracted from the \textsc{MOSFiT}. 
In order to account for the potential time delay between the core collapse and the SBO, i.e., the onset of the optical emission, this distribution was convolved with a top-hat distribution non-zero only for times $40\,\mathrm{s} \leqslant t \leqslant 2\times10^4\,\mathrm{s}$. 
The times are motivated from propagation times of the shock through the stellar envelope of blue super giants~\cite{Kistler:2012as}. 

\begin{figure}[htb]
    \begin{minipage}{0.6\textwidth}
    \centering
    \includegraphics[width=1\linewidth]{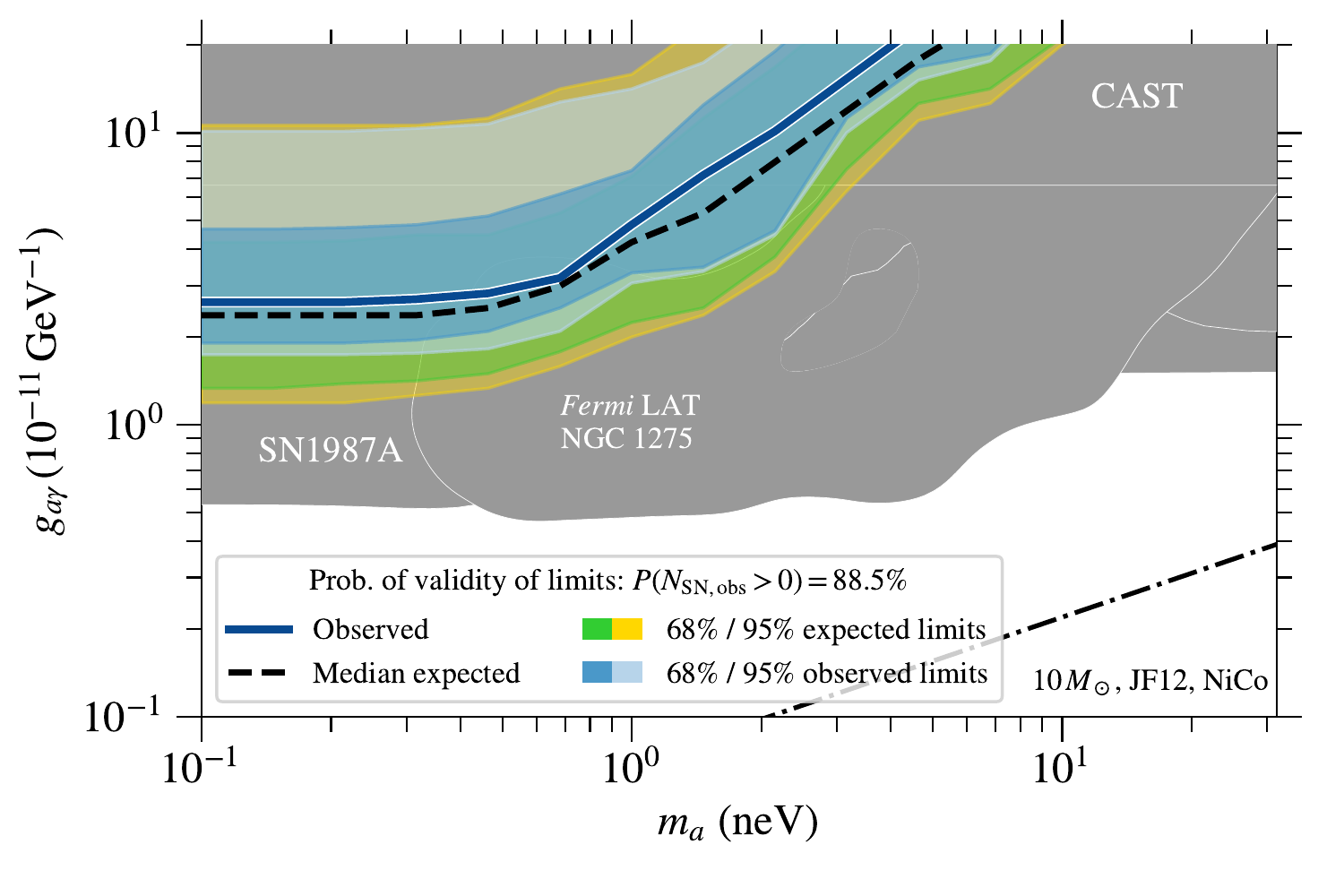}
    \end{minipage}
    \begin{minipage}{0.39\textwidth}
    \caption{
    Observed (blue shaded regions) and expected limits (green and yellow regions) on the ALP mass and coupling from the non-observation of a $\gamma$-ray burst from SNe 
    under the assumption that at least one SN was observed during the time of the core collapse.
    The median observed limit is shown as a blue solid line, whereas the median expected limit is shown as a dashed black line. 
    Grey shaded regions show already constrained parameters. Below the black dash-dotted line, ALPs could constitute the entire dark matter. Taken from Ref.~\cite{2020PhRvL.124w1101M} with the update from Ref.~\cite{2020PhRvL.125k9901M}.}
    \label{fig:sn-alps-results}
    \end{minipage}
\end{figure}

We searched for a $\gamma$-ray signal in the time window corresponding to $95\,\%$ quantile of the convolved distribution (centered on the peak of the distribution). Within this window, usually about one day wide, the search was conducted for each time interval of the satellite's orbit in which the position of the SN was in the LAT FOV.\footnote{Typically, a source is in the FOV for roughly 30 minutes during each 90 minute orbit.}
Time intervals far away from the best-fit time provided by \textsc{MOSFiT} were used as control regions.
No significant signal was found in the LAT data and we derived constraints on $g_{a\gamma}$ shown in Figure~\ref{fig:sn-alps-results} in blue.
The probability that at least one of the 20 SNe was in the FOV during the core collapse was estimated to be $P(N_\mathrm{SN, obs} \geqslant 1) \approx 89\,\%$ assuming the NiCo engine. 
The spread of the blue region in Figure~\ref{fig:sn-alps-results} denotes the statistical uncertainty as well as the uncertainty which SNe might have been actually observed by the LAT. 
The median limit is within a factor of 5 of the limits derived from the non-observation of a $\gamma$-ray signal of SN1987A~\cite{2015JCAP...02..006P}. 
Different assumptions on the structure of the Galactic magnetic field, the engine model (which affects the best estimate of the core-collapse time), and the progenitor mass (assuming 18 instead of 10\,$M_\odot$) all have a small impact on the final results. 
Furthermore, we estimated that with a sample of 40 SNe, there should be a 99\,\% chance that at least one core collapse occurred while in the LAT FOV. 
With the growing sample of SNe provided by currently operating optical surveys, it is therefore timely to revisit the SN sample to search for the ALP-induced $\gamma$-ray burst.

\section{Extending the SN sample with ZTF observations}
\label{sec:ztf-sample}
The SN discovery rate has accelerated in the past decade, owing to the number of wide-field robotic transient surveys. One of these optical surveys is the ZTF with a 47 square degree camera. 
The ZTF is an important  precursor to the upcoming Vera C. Rubin Observatory Legacy Survey of Space and Time (LSST; \cite{Ivezic_2019}) in the 2020s. 

We start by considering the recently published ZTF Bright Transient Survey (BST) Sample~\cite{Perley_2020}.
We limit the sample to SN Ib/c that have occurred within a redshift of $z \leqslant 0.02$ that corresponds to luminosity distance of $\rm d_L \leqslant 87$ Mpc in a flat $\Lambda$CDM cosmology with $H_0=70$ and $\Omega_m=0.3$. 
This leaves us with 21 objects from which we select 15 through  inspection by eye that have sufficient sampling and data close to the shock breakout in both $g$ and $r$ band\footnote{Our sample can be obtained through \url{https://sites.astro.caltech.edu/ztf/bts/explorer.php?f=s&subsample=ccsn&classexclude=II&quality=y&lcfig=y&ztflink=lasair&startz=0.0&endz=0.02&endpeakmag=19&format=table}}.
We extract the photometry together with the pre-explosion non-detection limits in $g$ and $r$ band from the ZTF BST database which we correct for the Milky Way dust extinction \cite{2011ApJ...737..103S}. Then, we fit the lightcurves with NiCo, SBO+NiCo, and exppow models \textsc{MOSFiT} using the same settings as in Ref.~ \cite{2020PhRvL.124w1101M}.
We include data points of the light curve within the interval $[-10, 50]$ days where $t=0$ corresponds to the first detection of the transient. 
The hydrogen column density, which determines the extinction in the host galaxy, is left as a free parameter. 
In addition to the explosion time $t_\mathrm{exp}$, In the NiCo model, the fit parameters are the mass of the ejecta, $M_\mathrm{ej}$, the fraction of Ni in the ejecta, $f_\mathrm{Ni}$, the velosity of the ejecta, $v_\mathrm{ej}$, the (high-energy) photon opacity $\kappa$ ($\kappa_\gamma$), the minimum temperature $T_\mathrm{min}$ for the photosphere, and an additional parameter $\sigma$ which can be interpreted as a model uncertainty and which brings the final $\chi^2$ value per degree of freedom of the fit close to one.
In the SBO+NiCo model an SBO component is added to the engine model, which has two additional parameters: a normalization $\L_\mathrm{scale}$ and $\alpha$ which determines the slope of the exponential decay \cite{2020PhRvL.124w1101M}. 
The exppow model has the free parameters $t_\mathrm{peak}$ for the time of the peak flux, a normalization parameter $\L_\mathrm{scale}$, as well as the shape parameters $\alpha$ and $\beta$ to describe the exponential rise and power-law decay.
For the parameters, we use flat priors in either linear or logarithmic space. 

\begin{figure*}
    \centering
    \includegraphics[width=.8\linewidth]{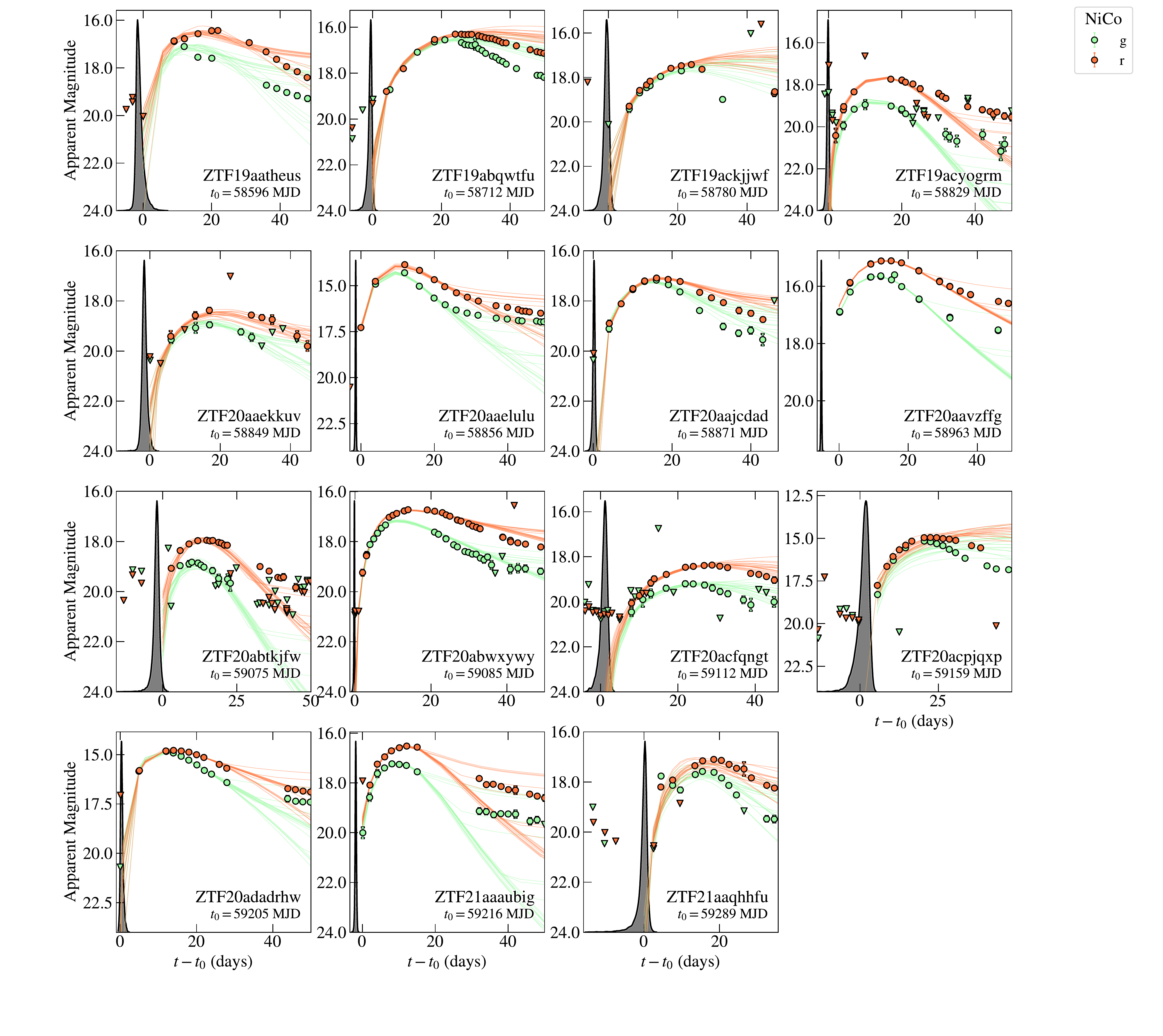}
    \caption{ZTF light curves for the $g$ and $r$ band together with model realizations drawn from the posterior distribution calculated from \textsc{MOSFiT} using the NiCo model. The posterior for the explosion time $t_\mathrm{exp}$ is shown as a black filled line.}
    \label{fig:nico}
\end{figure*}

\begin{figure*}
    \centering
    \includegraphics[width=.8\linewidth]{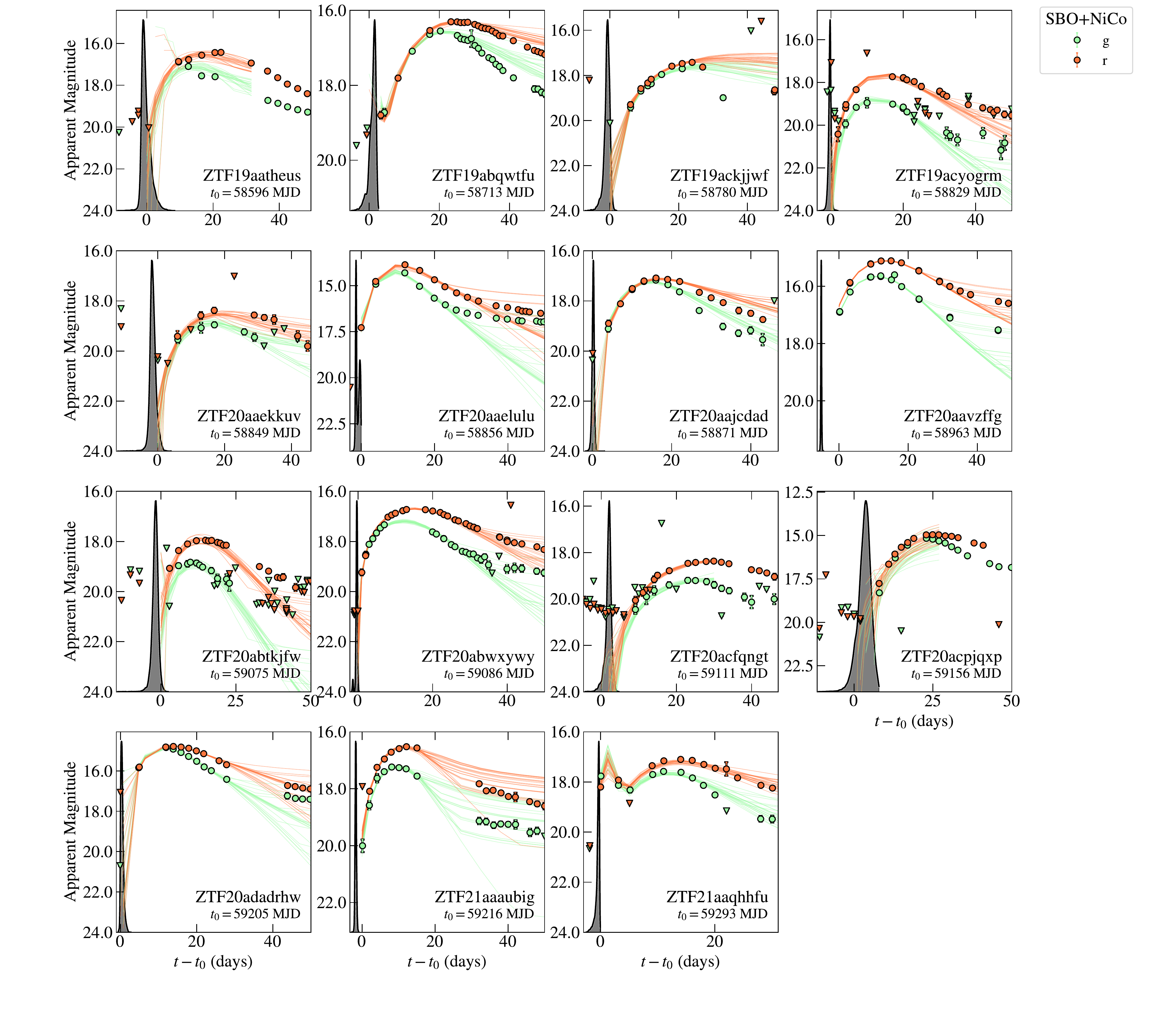}
    \caption{Same as Figure~\ref{fig:nico} for the SBO+NiCo model.}
    \label{fig:sbonico}
\end{figure*}

The best-fit parameters are listed in Tables~\ref{tab:mosfit}, \ref{tab:sbonico}, and \ref{tab:exppow} for the NiCo, SBO+Nico, and exppow models, respectively. 
The light curves are shown in Figures~\ref{fig:nico} and \ref{fig:sbonico} for the NiCo and SBO+NiCo models. 
The exppow model fails to describe the slow decrease of the flux for many light curves of our sample (also indicated by the higher best-fit values of $\sigma$), so we do not show the light curves for this model here. 
The NiCo and SBO+NiCo models on the other hand describe the light curves well, in particular the early parts up to 30 days after the first detection. 
The SBO+NiCo model achieves a better description for ZTF21aaqhhfu at early times since the additional SBO component is able to reproduce the flattening of the light curve right after the onset of the explosion (lowermost right panel of Fig.~\ref{fig:sbonico}). Consequently, the explosion time can be estimate with high accuracy and a statistical uncertainty of less than one day (see the last line of Table~\ref{tab:sbonico}).
Generally, $t_\mathrm{exp}$ is determined within an interval of 1-2 days, which is of the same order as for the sample used in Ref.~\cite{2020PhRvL.124w1101M}.

\begin{table*}
\centering
\begin{tiny}
\begin{tabular}{l|cccccccc}
\hline
\hline
Source name & $\log\, M_{\rm ej}\,(M_\odot)$ & $\log\, f_{\rm Ni}$ & $\kappa\,({\rm cm}^{2}\,{\rm g}^{-1})$ & $\log\, \kappa_\gamma\,({\rm cm}^{2}\,{\rm g}^{-1})$ & $\log\, v_{\rm ej}\,({\rm km\,s}^{-1})$ & $\log\, T_{\min}\,{\rm (K)}$ & $\log\, \sigma$ & $t_{\rm exp}\,{\rm (days)}$  \\
\hline
ZTF19aatheus & $0.49_{-0.24}^{+0.26}$ & $-1.00_{+0.34}^{-0.21}$ & $0.13_{-0.05}^{+0.04}$ & $1.83_{-1.74}^{+1.40}$ & $4.32_{-0.13}^{+0.12}$ & $3.66_{-0.33}^{+0.07}$ & $-0.74_{+0.11}^{-0.07}$ & $-10.23_{+1.20}^{-0.67}$\\
ZTF19abqwtfu & $0.08_{-0.14}^{+0.34}$ & $-0.57_{+0.21}^{-0.25}$ & $0.14_{-0.05}^{+0.04}$ & $1.53_{-1.54}^{+1.55}$ & $3.72_{-0.07}^{+0.07}$ & $3.59_{-0.36}^{+0.35}$ & $-1.11_{+0.15}^{-0.15}$ & $-4.72_{+0.43}^{-0.77}$\\
ZTF19ackjjwf & $0.56_{-0.42}^{+0.66}$ & $-1.25_{+0.34}^{-0.35}$ & $0.11_{-0.04}^{+0.05}$ & $0.67_{-1.13}^{+1.65}$ & $3.66_{-0.06}^{+0.33}$ & $3.70_{-0.42}^{+0.24}$ & $-2.01_{+0.43}^{-0.47}$ & $-6.56_{+0.68}^{-0.86}$\\
ZTF19acyogrm & $0.76_{-0.17}^{+0.19}$ & $-1.45_{+0.16}^{-0.19}$ & $0.11_{-0.04}^{+0.05}$ & $1.70_{-1.72}^{+1.45}$ & $4.49_{-0.05}^{+0.04}$ & $3.28_{-0.17}^{+0.17}$ & $-1.18_{+0.12}^{-0.11}$ & $-2.11_{+0.27}^{-0.34}$\\
ZTF20aaekkuv & $-0.18_{+0.32}^{-0.30}$ & $-1.10_{+0.42}^{-0.22}$ & $0.12_{-0.05}^{+0.05}$ & $1.63_{-1.62}^{+1.42}$ & $3.67_{-0.10}^{+0.08}$ & $3.45_{-0.27}^{+0.30}$ & $-2.01_{+0.55}^{-0.52}$ & $-7.51_{+0.71}^{-0.58}$\\
ZTF20aaelulu & $-0.12_{+0.23}^{-0.14}$ & $-0.78_{+0.15}^{-0.20}$ & $0.13_{-0.04}^{+0.05}$ & $1.48_{-1.53}^{+1.59}$ & $4.25_{-0.04}^{+0.04}$ & $3.35_{-0.21}^{+0.21}$ & $-1.12_{+0.13}^{-0.10}$ & $-1.42_{+0.09}^{-0.11}$\\
ZTF20aajcdad & $-0.28_{+0.20}^{-0.16}$ & $-0.62_{+0.16}^{-0.20}$ & $0.12_{-0.04}^{+0.05}$ & $2.26_{-1.82}^{+1.15}$ & $3.84_{-0.03}^{+0.03}$ & $3.49_{-0.29}^{+0.27}$ & $-2.12_{+0.55}^{-0.50}$ & $-3.83_{+0.25}^{-0.29}$\\
ZTF20aavzffg & $0.14_{-0.15}^{+0.17}$ & $-1.17_{+0.14}^{-0.18}$ & $0.12_{-0.04}^{+0.05}$ & $0.94_{-1.24}^{+1.78}$ & $3.93_{-0.01}^{+0.01}$ & $3.37_{-0.22}^{+0.20}$ & $-2.22_{+0.39}^{-0.39}$ & $-5.31_{+0.10}^{-0.10}$\\
ZTF20abtkjfw & $0.51_{-0.24}^{+0.24}$ & $-1.43_{+0.24}^{-0.21}$ & $0.11_{-0.03}^{+0.05}$ & $1.62_{-1.66}^{+1.52}$ & $4.26_{-0.11}^{+0.10}$ & $3.25_{-0.15}^{+0.15}$ & $-0.52_{+0.07}^{-0.05}$ & $-4.93_{+0.76}^{-0.99}$\\
ZTF20abwxywy & $-0.27_{+0.04}^{-0.04}$ & $-0.08_{+0.07}^{-0.26}$ & $0.05_{-0.00}^{+0.01}$ & $1.54_{-1.52}^{+1.53}$ & $4.08_{-0.02}^{+0.02}$ & $3.44_{-0.27}^{+0.26}$ & $-2.08_{+0.49}^{-0.53}$ & $-2.17_{+0.07}^{-0.10}$\\
ZTF20acfqngt & $0.58_{-0.23}^{+0.24}$ & $-1.48_{+0.20}^{-0.18}$ & $0.11_{-0.04}^{+0.04}$ & $1.96_{-1.75}^{+1.29}$ & $3.74_{-0.06}^{+0.05}$ & $3.45_{-0.25}^{+0.25}$ & $-2.20_{+0.43}^{-0.44}$ & $-7.02_{+0.72}^{-1.05}$\\
ZTF20acpjqxp & $1.22_{-0.45}^{+0.30}$ & $-1.21_{+0.64}^{-0.39}$ & $0.13_{-0.05}^{+0.04}$ & $1.45_{-1.55}^{+1.55}$ & $3.98_{-0.40}^{+0.38}$ & $3.65_{-0.39}^{+0.25}$ & $-0.12_{+0.07}^{-0.04}$ & $-4.31_{+1.38}^{-2.41}$\\
ZTF20adadrhw & $-0.51_{+0.22}^{-0.12}$ & $-0.29_{+0.12}^{-0.24}$ & $0.14_{-0.05}^{+0.04}$ & $1.75_{-2.71}^{+1.34}$ & $4.04_{-0.02}^{+0.02}$ & $3.37_{-0.22}^{+0.22}$ & $-1.51_{+0.14}^{-0.17}$ & $-4.51_{+0.37}^{-0.27}$\\
ZTF21aaaubig & $0.20_{-0.16}^{+0.18}$ & $-1.46_{+0.15}^{-0.17}$ & $0.11_{-0.04}^{+0.05}$ & $1.83_{-1.62}^{+1.41}$ & $4.15_{-0.02}^{+0.02}$ & $3.36_{-0.22}^{+0.22}$ & $-1.77_{+0.35}^{-0.45}$ & $-1.94_{+0.17}^{-0.17}$\\
ZTF21aaqhhfu & $-0.24_{+0.65}^{-0.24}$ & $-1.48_{+0.33}^{-0.43}$ & $0.12_{-0.05}^{+0.05}$ & $1.48_{-1.53}^{+1.55}$ & $3.52_{-0.16}^{+0.35}$ & $3.57_{-0.34}^{+0.32}$ & $-0.36_{+0.09}^{-0.06}$ & $-4.51_{+0.65}^{-1.58}$\\

\hline
\end{tabular}
\end{tiny}
\caption{\label{tab:mosfit} Best-fit parameters for the NiCo model.}
\end{table*}

\begin{table*}
\centering
\begin{tiny}
\begin{tabular}{l|cccccccccc}
\hline
\hline
Source name & $\log\, M_{\rm ej}\,(M_\odot)$ & $f_{\rm Ni}$ & $\kappa\,({\rm cm}^{2}\,{\rm g}^{-1})$ & $\log\, \kappa_\gamma\,({\rm cm}^{2}\,{\rm g}^{-1})$ & $\log\, v_{\rm ej}\,({\rm km\,s}^{-1})$ & $\log\, T_{\min}\,{\rm (K)}$ & $\alpha$ & $\log\, \L_{\rm scale}$ & $\log\, \sigma$ & $t_{\rm exp}\,{\rm (days)}$  \\
\hline
ZTF19aatheus & $0.39_{-0.22}^{+0.22}$ & $0.13_{-0.05}^{+0.48}$ & $0.13_{-0.05}^{+0.04}$ & $1.87_{-1.51}^{+1.35}$ & $4.22_{-0.17}^{+0.15}$ & $3.65_{-0.34}^{+0.10}$ & $19.12_{-10.04}^{+6.72}$ & $43.34_{-5.03}^{+5.67}$ & $-0.68_{+0.11}^{-0.09}$ & $-10.06_{+1.63}^{-0.84}$\\
ZTF19abqwtfu & $-0.13_{+0.12}^{-0.09}$ & $0.51_{-0.12}^{+0.16}$ & $0.15_{-0.04}^{+0.03}$ & $0.32_{-0.95}^{+1.38}$ & $3.78_{-0.05}^{+0.04}$ & $3.50_{-0.31}^{+0.31}$ & $8.56_{-1.77}^{+2.60}$ & $47.92_{-1.09}^{+1.60}$ & $-2.03_{+0.53}^{-0.49}$ & $-2.42_{+0.84}^{-1.47}$\\
ZTF19ackjjwf & $-0.16_{-0.10}^{+0.28}$ & $0.29_{-0.16}^{+0.26}$ & $0.12_{-0.03}^{+0.03}$ & $0.30_{-0.86}^{+1.54}$ & $3.56_{-0.07}^{+0.07}$ & $3.73_{-0.28}^{+0.22}$ & $19.32_{-7.20}^{+6.15}$ & $48.01_{-5.53}^{+3.38}$ & $-2.09_{+0.47}^{-0.46}$ & $-6.75_{+0.65}^{-0.86}$\\
ZTF19acyogrm & $0.20_{-0.11}^{+0.11}$ & $0.27_{-0.08}^{+0.16}$ & $0.16_{-0.04}^{+0.02}$ & $0.68_{-1.11}^{+1.59}$ & $4.39_{-0.05}^{+0.04}$ & $3.35_{-0.21}^{+0.20}$ & $16.48_{-8.13}^{+8.09}$ & $43.10_{-4.86}^{+5.23}$ & $-1.16_{+0.11}^{-0.10}$ & $-2.24_{+0.29}^{-0.36}$\\
ZTF20aaekkuv & $-0.43_{+0.26}^{-0.24}$ & $0.15_{-0.07}^{+0.23}$ & $0.13_{-0.04}^{+0.04}$ & $1.48_{-1.37}^{+1.41}$ & $3.59_{-0.08}^{+0.08}$ & $3.68_{-0.33}^{+0.22}$ & $20.73_{-7.62}^{+5.69}$ & $42.91_{-4.80}^{+5.23}$ & $-1.63_{+0.44}^{-0.54}$ & $-7.42_{+0.90}^{-0.68}$\\
ZTF20aaelulu & $-0.32_{+0.16}^{-0.14}$ & $0.27_{-0.08}^{+0.10}$ & $0.17_{-0.03}^{+0.02}$ & $1.22_{-1.37}^{+1.66}$ & $4.25_{-0.04}^{+0.03}$ & $3.24_{-0.14}^{+0.17}$ & $21.05_{-7.27}^{+5.20}$ & $47.81_{-6.15}^{+2.55}$ & $-1.09_{+0.12}^{-0.08}$ & $-1.25_{+0.96}^{-0.24}$\\
ZTF20aajcdad & $-0.38_{+0.16}^{-0.12}$ & $0.29_{-0.09}^{+0.11}$ & $0.15_{-0.05}^{+0.03}$ & $1.72_{-1.54}^{+1.38}$ & $3.83_{-0.04}^{+0.03}$ & $3.50_{-0.29}^{+0.26}$ & $17.07_{-9.56}^{+8.36}$ & $41.02_{-3.77}^{+5.55}$ & $-1.59_{+0.27}^{-0.31}$ & $-3.82_{+0.23}^{-0.28}$\\
ZTF20aavzffg & $0.02_{-0.06}^{+0.09}$ & $0.09_{-0.02}^{+0.01}$ & $0.16_{-0.03}^{+0.02}$ & $1.82_{-1.71}^{+1.43}$ & $3.93_{-0.01}^{+0.01}$ & $3.36_{-0.22}^{+0.21}$ & $17.85_{-8.90}^{+7.34}$ & $43.09_{-5.15}^{+6.91}$ & $-2.19_{+0.41}^{-0.41}$ & $-5.32_{+0.10}^{-0.11}$\\
ZTF20abtkjfw & $0.32_{-0.19}^{+0.18}$ & $0.05_{-0.02}^{+0.06}$ & $0.16_{-0.06}^{+0.03}$ & $1.58_{-1.62}^{+1.53}$ & $4.23_{-0.13}^{+0.10}$ & $3.27_{-0.15}^{+0.14}$ & $16.09_{-8.83}^{+8.55}$ & $44.43_{-6.19}^{+7.22}$ & $-0.47_{+0.07}^{-0.06}$ & $-4.69_{+0.80}^{-0.94}$\\
ZTF20abwxywy & $-0.26_{+0.25}^{-0.08}$ & $0.26_{-0.12}^{+0.17}$ & $0.15_{-0.03}^{+0.03}$ & $2.76_{-1.24}^{+0.81}$ & $4.13_{-0.04}^{+0.03}$ & $3.37_{-0.22}^{+0.25}$ & $1.62_{-0.52}^{+0.55}$ & $43.43_{-0.52}^{+0.44}$ & $-1.78_{+0.29}^{-0.25}$ & $-1.34_{+0.17}^{-0.21}$\\
ZTF20acfqngt & $0.00_{-0.18}^{+0.31}$ & $0.17_{-0.12}^{+0.29}$ & $0.12_{-0.02}^{+0.03}$ & $2.69_{-0.86}^{+0.67}$ & $3.65_{-0.07}^{+0.08}$ & $3.58_{-0.21}^{+0.24}$ & $15.45_{-5.47}^{+6.74}$ & $47.01_{-3.15}^{+3.06}$ & $-2.10_{+0.42}^{-0.47}$ & $-7.08_{+0.55}^{-0.80}$\\
ZTF20acpjqxp & $0.97_{-0.60}^{+0.43}$ & $0.42_{-0.27}^{+0.32}$ & $0.14_{-0.05}^{+0.04}$ & $1.58_{-1.50}^{+1.55}$ & $3.67_{-0.35}^{+0.35}$ & $3.75_{-0.41}^{+0.39}$ & $15.39_{-9.85}^{+9.10}$ & $45.10_{-6.07}^{+5.11}$ & $-0.11_{+0.06}^{-0.05}$ & $-4.62_{+1.77}^{-2.31}$\\
ZTF20adadrhw & $-0.60_{+0.10}^{-0.09}$ & $0.62_{-0.13}^{+0.13}$ & $0.17_{-0.04}^{+0.02}$ & $-0.98_{+2.16}^{+0.04}$ & $4.04_{-0.02}^{+0.02}$ & $3.39_{-0.23}^{+0.21}$ & $18.40_{-9.20}^{+7.45}$ & $43.60_{-5.33}^{+6.55}$ & $-1.51_{+0.15}^{-0.20}$ & $-4.43_{+0.39}^{-0.27}$\\
ZTF21aaaubig & $0.09_{-0.08}^{+0.09}$ & $0.05_{-0.01}^{+0.02}$ & $0.11_{-0.02}^{+0.02}$ & $0.37_{-0.58}^{+0.72}$ & $4.13_{-0.04}^{+0.04}$ & $3.63_{-0.11}^{+0.07}$ & $4.89_{-2.70}^{+5.17}$ & $37.89_{-1.83}^{+2.02}$ & $-1.29_{+0.11}^{-0.12}$ & $-1.84_{+0.17}^{-0.17}$\\
ZTF21aaqhhfu & $-0.71_{+0.16}^{-0.13}$ & $0.10_{-0.04}^{+0.05}$ & $0.16_{-0.04}^{+0.03}$ & $1.52_{-1.43}^{+1.39}$ & $3.63_{-0.07}^{+0.06}$ & $3.44_{-0.25}^{+0.24}$ & $10.56_{-1.83}^{+1.54}$ & $49.46_{-1.46}^{+1.29}$ & $-0.83_{+0.12}^{-0.10}$ & $-0.51_{+0.23}^{-0.59}$\\

\hline
\end{tabular}
\end{tiny}
\caption{\label{tab:sbonico} Best-fit parameters for the SBO+NiCo model.}
\end{table*}

\begin{table*}
\centering
\begin{tiny}
\begin{tabular}{l|cccccccccc}
\hline
\hline
Source name & $\alpha$ & $\beta$ & $\log\, t_{\rm peak}$ & $\log\, \L_{\rm scale}$ & $\kappa\,({\rm cm}^{2}\,{\rm g}^{-1})$ & $\log\, \kappa_\gamma\,({\rm cm}^{2}\,{\rm g}^{-1})$ & $\log\, M_{\rm ej}\,(M_\odot)$ & $\log\, v_{\rm ej}\,({\rm km\,s}^{-1})$ & $\log\, \sigma$ & $t_{\rm exp}\,{\rm (days)}$  \\
\hline
ZTF19aatheus & $5.68_{-1.86}^{+1.94}$ & $4.23_{-1.39}^{+2.11}$ & $-0.16_{-0.31}^{+0.70}$ & $46.58_{-1.48}^{+0.88}$ & $0.15_{-0.03}^{+0.03}$ & $0.70_{-0.83}^{+1.29}$ & $1.45_{-0.24}^{+0.16}$ & $4.01_{-0.09}^{+0.10}$ & $-0.78_{+0.10}^{-0.07}$ & $-7.95_{+1.71}^{-1.34}$\\
ZTF19abqwtfu & $5.87_{-2.01}^{+1.98}$ & $1.34_{-0.54}^{+0.76}$ & $0.51_{-0.98}^{+0.27}$ & $45.44_{-0.84}^{+0.89}$ & $0.07_{-0.02}^{+0.03}$ & $1.45_{-1.35}^{+1.11}$ & $-0.26_{-0.23}^{+0.44}$ & $3.26_{-0.10}^{+0.13}$ & $-1.01_{+0.13}^{-0.10}$ & $-5.23_{+2.10}^{-1.68}$\\
ZTF19ackjjwf & $7.04_{-2.55}^{+1.47}$ & $0.88_{-0.56}^{+0.63}$ & $0.32_{-0.30}^{+0.31}$ & $43.62_{-0.68}^{+0.69}$ & $0.09_{-0.02}^{+0.03}$ & $3.43_{-0.53}^{+0.35}$ & $0.14_{-0.25}^{+0.45}$ & $3.37_{-0.12}^{+0.09}$ & $-2.32_{+0.32}^{-0.39}$ & $-7.03_{+1.36}^{-2.00}$\\
ZTF19acyogrm & $3.71_{-1.47}^{+1.37}$ & $5.00_{-1.77}^{+1.31}$ & $-0.28_{+0.58}^{-0.25}$ & $47.15_{-1.23}^{+0.51}$ & $0.14_{-0.04}^{+0.03}$ & $3.18_{-0.77}^{+0.54}$ & $0.26_{-0.18}^{+0.29}$ & $4.16_{-0.06}^{+0.07}$ & $-1.09_{+0.11}^{-0.10}$ & $-0.68_{+0.20}^{-0.49}$\\
ZTF20aaelulu & $1.69_{-0.73}^{+1.25}$ & $2.34_{-0.59}^{+1.01}$ & $0.88_{-0.23}^{+0.24}$ & $43.57_{-0.42}^{+1.21}$ & $0.07_{-0.01}^{+0.04}$ & $0.86_{-1.23}^{+0.98}$ & $0.28_{-0.17}^{+0.21}$ & $4.11_{-0.12}^{+0.07}$ & $-0.93_{+0.19}^{-0.10}$ & $-0.83_{+0.20}^{-0.35}$\\
ZTF20aaekkuv & $7.13_{-1.45}^{+1.56}$ & $2.24_{-0.49}^{+5.54}$ & $-0.87_{+1.68}^{-0.57}$ & $46.54_{-1.81}^{+1.01}$ & $0.17_{-0.07}^{+0.02}$ & $1.72_{-1.52}^{+1.16}$ & $0.21_{-0.38}^{+0.42}$ & $3.42_{-0.21}^{+0.10}$ & $-2.04_{+0.48}^{-0.51}$ & $-3.94_{+1.05}^{-1.49}$\\
ZTF20aajcdad & $8.68_{-1.37}^{+0.85}$ & $7.15_{-1.26}^{+1.05}$ & $1.25_{-0.15}^{+0.15}$ & $44.10_{-0.35}^{+0.32}$ & $0.12_{-0.04}^{+0.04}$ & $1.32_{-1.44}^{+1.60}$ & $-0.20_{+0.24}^{-0.18}$ & $3.66_{-0.05}^{+0.04}$ & $-1.38_{+0.25}^{-0.22}$ & $-4.95_{+0.82}^{-0.83}$\\
ZTF20aavzffg & $6.75_{-2.94}^{+1.96}$ & $0.99_{-0.13}^{+0.13}$ & $-1.72_{+0.25}^{-0.17}$ & $45.10_{-0.37}^{+0.41}$ & $0.09_{-0.03}^{+0.04}$ & $0.99_{-1.36}^{+1.66}$ & $0.23_{-0.16}^{+0.16}$ & $3.86_{-0.02}^{+0.02}$ & $-1.37_{+0.12}^{-0.09}$ & $-2.86_{+0.36}^{-0.37}$\\
ZTF20abtkjfw & $4.00_{-1.83}^{+1.86}$ & $2.21_{-0.79}^{+1.31}$ & $0.48_{-0.28}^{+0.52}$ & $43.98_{-0.64}^{+0.62}$ & $0.13_{-0.04}^{+0.04}$ & $-0.60_{-0.26}^{+2.39}$ & $0.76_{-0.41}^{+0.24}$ & $4.18_{-0.16}^{+0.13}$ & $-0.52_{+0.08}^{-0.06}$ & $-5.40_{+2.41}^{-4.98}$\\
ZTF20abwxywy & $1.57_{-0.97}^{+1.30}$ & $1.21_{-0.10}^{+0.09}$ & $-0.90_{+0.35}^{-0.47}$ & $45.09_{-0.45}^{+0.48}$ & $0.16_{-0.03}^{+0.02}$ & $-0.58_{-0.30}^{+0.78}$ & $0.36_{-0.11}^{+0.12}$ & $4.10_{-0.03}^{+0.02}$ & $-2.43_{+0.43}^{-0.35}$ & $-0.97_{+0.15}^{-0.21}$\\
ZTF20acfqngt & $4.97_{-1.96}^{+3.89}$ & $3.04_{-1.60}^{+3.68}$ & $1.01_{-0.80}^{+0.33}$ & $44.46_{-1.00}^{+0.68}$ & $0.12_{-0.05}^{+0.04}$ & $2.03_{-1.57}^{+1.20}$ & $0.45_{-0.37}^{+0.40}$ & $3.37_{-0.08}^{+0.06}$ & $-1.91_{+0.25}^{-0.26}$ & $-10.03_{+5.05}^{-4.39}$\\
ZTF20acpjqxp & $5.74_{-2.71}^{+2.48}$ & $1.68_{-1.31}^{+2.47}$ & $1.20_{-1.42}^{+0.48}$ & $44.34_{-0.75}^{+1.10}$ & $0.14_{-0.05}^{+0.04}$ & $1.53_{-1.43}^{+1.53}$ & $0.92_{-0.69}^{+0.47}$ & $3.74_{-0.36}^{+0.31}$ & $-0.15_{+0.06}^{-0.04}$ & $-9.76_{+5.03}^{-5.27}$\\
ZTF20adadrhw & $3.71_{-1.42}^{+1.82}$ & $6.41_{-1.52}^{+1.72}$ & $0.90_{-0.61}^{+0.59}$ & $41.86_{-1.72}^{+1.92}$ & $0.08_{-0.02}^{+0.06}$ & $3.22_{-0.69}^{+0.52}$ & $1.11_{-0.41}^{+0.28}$ & $3.88_{-0.07}^{+0.06}$ & $-1.02_{+0.09}^{-0.06}$ & $-0.79_{+0.20}^{-0.40}$\\
ZTF21aaaubig & $7.84_{-1.80}^{+1.35}$ & $6.20_{-1.25}^{+1.16}$ & $0.88_{-0.10}^{+0.09}$ & $44.51_{-0.50}^{+0.50}$ & $0.16_{-0.04}^{+0.03}$ & $2.36_{-1.61}^{+1.10}$ & $0.62_{-0.25}^{+0.36}$ & $3.90_{-0.06}^{+0.05}$ & $-1.29_{+0.15}^{-0.18}$ & $-2.42_{+0.64}^{-0.70}$\\
ZTF21aaqhhfu & $5.69_{-3.56}^{+3.19}$ & $4.90_{-1.51}^{+2.05}$ & $0.16_{-0.44}^{+0.99}$ & $45.82_{-3.12}^{+1.44}$ & $0.15_{-0.04}^{+0.03}$ & $2.23_{-2.15}^{+1.25}$ & $0.73_{-0.41}^{+0.46}$ & $3.22_{-0.11}^{+0.15}$ & $-0.42_{+0.08}^{-0.06}$ & $-2.08_{+0.40}^{-0.46}$\\

\hline
\end{tabular}
\end{tiny}
\caption{\label{tab:exppow} Best-fit parameters for the exppow model.}
\end{table*}

\section{Conclusions}
\label{sec:conclusions}
A $\gamma$-ray burst signal from a core-collapse SN at energies of tens of MeV coincident with the neutrino burst is a smoking-gun signature for the existence of ALPs. 
Standard processes should lead to $\gamma$-ray emission only after some delay and the emission is expected to peak at lower energies~\cite{hjorth_bloom_2012}. 
As current neutrino telescopes are not sensitive enough to detect SN neutrinos beyond the Andromeda Galaxy, the time at which one has to search for an ALP-induced burst must be estimated from optical SN observations. 
In Ref.~\cite{2020PhRvL.124w1101M}, we have carried out this search using a sample of 20 well-studied SNe of type Ib/c.
The results are promising: with a probability of almost 90\,\% at least one core collapse occurred while in the FOV of the LAT.
Assuming that this is the case, the non-observation of a burst leads to constraints on the photon-ALP coupling of $g_{a\gamma} < 2.6\times10^{-11}\,\mathrm{GeV}^{-1}$ for $m_a \lesssim 0.3\,$neV. 

In order to increase the probability that at least one SN has been in the FOV, one has to consider a larger sample of SNe with well-sampled light curves. 
We have taken a first look of SNe of type Ib/c with redshifts $z < 0.02$ observed with ZTF. 
For our considered selection of 15 SNe, we are able to determine the explosion time within a time interval of 1-2 days (statistical uncertainties only), which is similar to the accuracy of the original sample considered in Ref.~\cite{2020PhRvL.124w1101M}.
Therefore, these SNe should be excellent candidates for a follow-up analysis of LAT data. 
With distances between $0.0052 \leqslant z \leqslant 0.02$ (corresponding to luminosity distances between $\sim 23$ and $ \sim88$\,Mpc) the SNe in the ZTF sample could lead to comparable constraints as in the original sample and stacking the observations should further tighten the limits on the ALP coupling. 

Assuming the SBO+NiCo model, 6 out of 15 SNe appear to have been observed within less than three days after the SBO in roughly two years of observation time. This number is within a factor of $\sim 2$ of our original estimate that ZTF should roughly observe 6 Ib/c SNe  per year within the first three days after the explosion~\cite{2020PhRvL.124w1101M}. 
With the upcoming Rubin Observatory, which could detect up to 20 SNe of type Ib/c up to redshift $0.02$ within one day of the explosion, the SN sample is expected to grow even more rapidly in the near future. 
Lastly, we would like to point out that ZTF19abqwtfu and ZTF20aaelulu were within the field of view of the TESS satellite during their peak emission. Since the TESS satellite continuosly monitors a given sky region for 28 consecutive days, the inclusion of TESS data in the \textsc{MOSFiT} parameter estimation  could tremendously improve the accuracy with which $t_\mathrm{exp}$ is determined as demonstrated in Ref.~\cite{2021MNRAS.500.5639V}. 

\begin{acknowledgments}
M.  M.  acknowledges  support from the European Research Council (ERC) under the European Union’s Horizon 2020 research and innovation program Grant agreement No. 948689 (AxionDM).
The \textit{Fermi}-LAT Collaboration acknowledges support for LAT development, operation and data analysis from NASA and DOE (United States), CEA/Irfu and IN2P3/CNRS (France), ASI and INFN (Italy), MEXT, KEK, and JAXA (Japan), and the K.A.~Wallenberg Foundation, the Swedish Research Council and the National Space Board (Sweden). Science analysis support in the operations phase from INAF (Italy) and CNES (France) is also gratefully acknowledged. This work performed in part under DOE Contract DE-AC02-76SF00515.
\end{acknowledgments}

\bibliographystyle{jhep}
\bibliography{mainbib}

%
%
%

\end{document}